\documentstyle[aps,prl,preprint]{revtex}

\newcommand{\be}{\begin{equation}}
\newcommand{\ee}{\end{equation}}
\newcommand{\ber}{\begin{eqnarray}}
\newcommand{\eer}{\end{eqnarray}}

\input epsfig.sty

\begin{document}
\tighten
\draft

\title{Weak Chaos in a Quantum Kepler Problem.}

\author{B. L. Altshuler$^a$ and L. S. Levitov$^{b}$}
\address{(a) NEC Research Institute and Princeton University;\\
(b) MIT, 12-112,
77 Massachusetts Ave., Cambridge, MA 02139}
\maketitle

\begin{abstract}
Transition from regular to chaotic dynamics in a crystal made of
singular scatterers $U(r)=\lambda |r|^{-\sigma}$ can be reached
by varying either $\sigma$ or $\lambda$. We map the problem to a
localization problem, and find that in all space dimensions the
transition occurs at $\sigma=1$, i.e., Coulomb potential has
marginal singularity. We study the critical line $\sigma=1$ by
means of a renormalization group technique, and describe
universality classes of this new transition. An RG equation is
written in the basis of states localized in momentum space. The
RG flow evolves the distribution of coupling parameters to a
universal stationary distribution. Analytic properties of the RG
equation are similar to that of Boltzmann kinetic equation: the
RG dynamics has integrals of motion and obeys an $H-$theorem.
The RG results for $\sigma=1$ are used to derive scaling laws
for transport and to calculate critical exponents.

\end{abstract}




\section{Introduction}
In the theory of localization to which I.~M.~Lifshits made  many
seminal contributions\cite{Lifshitz}, one is interested in the
effect of random potential on the quantum-mechanical
wavefunction. When the randomness is weak, the wavefunctions are
extended throughout the whole system, whereas at sufficiently high
disorder all wavefunctions become localized\cite{Anderson}. The transition
between localized and extended states for many years has been at
the center of interest in condensed matter physics\cite{Ziman}.  Originally,
the  concept of localization was invented in order to understand
properties of disordered metals and doped  semiconductors.  More
recently,  it  became  clear  that  the  ideas  and the language
developed in the  localization  theory  are  useful  in  a  much
broader  context.  One  of the areas of theoretical physics that
has numerous links  with  the  localization  theory  is  Quantum
Chaos.  In  this article we consider quantum chaos in a periodic
Kepler problem, and solve this problem by using tools  from  the
localization  theory. The relation between the Quantum Chaos and
Localization theories  appears  to  be  quite  generic,  and  we
believe  that  it  will  help  to understand the transition from
integrable to chaotic behavior  in  a  broad  class  of  quantum
systems.

Traditionally,  the  field  of  Quantum  Chaos  deals  with  the
relation  between  classical  and  quantum  dynamics  in chaotic
systems. One is interested in the features of classical dynamics
that are manifested in quantum  dynamics,  such  as  periodic
orbits,  ergodicity,  etc.  Another  group of questions concerns
the   transition  from  integrable  and  chaotic  quantum
dynamics. Already in the classical case this transition  appears
to  be  a difficult  problem, although tractable within the KAM theory\cite{KAM,Gutzwiller}.
On the other hand, the transition in the quantum case, apparently,
is not quite so well understood. For the reader's sake, below we review some
basic facts.

It is known that classical Hamiltonian dynamics
can be either integrable or chaotic. Integrability means that
all variables can be separated, which leads to quasiperiodic
motion, concentrated in a small part of the phase space. In
contrast, the chaotic (nonintegrable) dynamics is ergodic,
i.e., the system can be found arbitrarily close to a given point
in the phase space.

One of the most challenging problems
in the theory of classical dynamical systems is the transition
from integrable to chaotic behavior when an integrable system is
perturbed, and the perturbation violates the integrability.
There exist two scenarios of this transition. In the case of a
singular perturbation, even a perturbation of an arbitrarily
small amplitude can drive the system into chaos. A smooth
perturbation acts in a different way: according to the seminal
KAM scenario (see~\cite{KAM,Gutzwiller}) the motion remains non-ergodic
under a smooth perturbation if the latter is weak.

A good example of a singular classical Hamiltonian dynamics is
the motion in a periodic lattice of Coulomb
scatterers\cite{Knauf}. The dynamics in this problem can be
mapped onto the problem of geodesics on a negative curvature
compact space, which in the dynamical systems classification is
known to be of the most unpredictable, and yet deterministic
kind. The degree of ergodicity one has in this problem is
expressed mathematically by so-called Bernoulli symbolic
dynamics scheme. This dynamics leads to very fast mixing in the
phase space quantified by Kolmogorov entropy. Another
interesting example of a similar kind is the anisotropic Kepler
problem\cite{Gutzwiller1}: a particle with anisotropic mass
tensor moving in the Coulomb potential. It was shown by
Gutzwiller that the existence of the trajectories that fall on
the center breaks integrability at any potential strength.
Qualitative features of the dynamics are similar to that of the
periodic system of Coulomb scatterers, although the Bernoulli
scheme is constructed in a different way\cite{Gutzwiller1}.

In  the  quantum problem, a lot of interest recently was focused
on the semiclassical theory of quantum billiards. The  billirads
represent  one  of  the simplest problems in which one can study
the relation between classical and  quantum  chaos  (see  review
in\cite{Smilansky}).  The quantum problem of a periodic array of
Coulomb  scatterers was considered by Gutzwiller who developed a
scattering mode  representation\cite{quantumSinai}  well  suited
for treating the semiclassical limit.

In this article we consider the integrability to chaos
transition in {\it quantum} systems, focusing on the case of
singular perturbation. The usual approach is to look for
critical strength of the perturbation. Instead of that, however,
we found it more natural and interesting to add another
dimension to the problem, and to characterize the integrable to
non-integrable transition as a function of the degree of
singularity of the perturbation. It turns out that if the
perturbation is both not strong and not singular enough the
quantum behavior, namely the spectral statistics, will remain
similar to the integrable one. On the other hand, when either
the perturbation itself or its singularity is sufficiently
strong, the behavior is chaotic. We also find an intermediate
behavior for weak perturbations with the singularity taking
certain critical value. This behavior can be called quasichaos,
or {\it weak chaos}.

The problem we will be interested in is the motion of a quantum
particle in a three-dimensional crystal of scatterers with
Coulomb cores. It will be assumed that the particle velocity is
so high that the wavelength is much smaller than the crystal
period. This problem describes transmission of fast charged
particles through a crystal, assuming the particles energy is so high that
scattering by the atomic core potential is important. (We ignore complications
arising from inelastic scattering.) The relation of our problem
to the real situation corresponding to
experiments on channeling particles in crystals will be
discussed at the end of the paper in Sec.~\ref{Channeling}.

Usually, for a particle moving in a periodic potential there is
a notion of a Bloch state, i.e., the particle wavefunction can
be represented as a superposition of few plane waves. However,
when the scattering by the cores is strong the number of plane
waves forming each Bloch state becomes very large, and the
wavevectors are distributed roughly uniformly over all
directions in the momentum space. This means that
the particle is not moving along a straight line, but rather is
diffusing. These two regimes, ballistic motion roughly along a
straight line and diffusive motion with an irregular change of
direction, correspond to the Bloch states {\it localized} or
{\it delocalized} in momentum space.

The idea of our approach is the following. We treat the core
potential as a perturbation, and write the problem using plane waves as a basis.
In this basis,
the main part of Hamiltonian, the kinetic
energy, is diagonal. Since the scattering potential is periodic,
the wavevectors of relevant plane waves form a lattice in
momentum space. In the chosen basis we have a tight-binding
localization problem: the potential matrix elements define
hopping between different lattice sites, and the kinetic energy
can be interpreted as on-site energy.

For a non-singular core potential one effectively has only
short-range hopping in the localization problem in momentum
space. In this case, according to the conventional theory of
localization, all states remain localized if the potential
strength is below some threshold. On the other hand, if the core
potential is singular, its Fourier components decay slowly and
one has to deal with a localization problem with long-range
hopping.
We will see that long-range hopping canm delocalize the quantum states
even at arbitrarily weak potential, if the potential singularity is
sufficiently strong.

The states localized in momentum space correspond to
quasiperiodic motion in real space, i.e., to almost integrable
dynamics, whereas the states delocalized in the momentum space
correspond to chaotic dynamics in real space. Generally, the
transition from one regime to the other can be reached in two
different ways: either by increasing the strength of the core
potential, or by varying the power law of the potential
singularity at the origin. Below we consider the situation when
the potential is weak: $Ze^2/\hbar v < 1$, where $v$ is the
particle velocity, and $Z$ is nuclear charge. In this case,
naively, the scattering is weak and thus the Bloch states
consist only of few plane waves. However, we will see below that
because of the $1/r-$singularity of the Coulomb potential the
condition of weak scattering is more stringent: $Ze^2/\hbar v
\ln p a_B/\hbar < 1$, where $p$ is the particle momentum, and
$a_B$ is Bohr's radius of the cores. We will be interested in
the situation when
  \be
Ze^2/\hbar v  < 1\ ,\qquad {\rm and}\qquad
Ze^2/\hbar v\ \ln (p a_B/\hbar)\ \gg 1\ ,
  \ee
In this case the scattering is strong, there is no ballistic
motion, and the Bloch states consist of many plane waves.

However, in this case the motion is not ergodic because the
number of plane waves forming the Bloch state is much less than
the total number of plane waves with appropriate energy. The
reason is, as we will see below, that this problem falls right
on the critical line of a localization transition. To study this
transition we introduce a more general model with core
potentials having an arbitrary power law singularity (see
Sec.~\ref{Model}), and consider the transition from localized to
extended states in momentum space resulting  from  changing  the
power  law.  The interesting thing about this transition is that
it occurs even in  the  weak  coupling  limit,  whereas  in  the
conventional  Anderson  localization  problem  with  short-range
hopping all states are localized at weak hopping. As  a  result,
the  nature  of  the  transition  is  quite  different  from the
conventional picture accepted  for  short-range  hopping,  which
leads to a completely different critical behavior.

After introducing the model in Sec.~\ref{Model}, and discussing
the mapping to a localization problem in Sec.~\ref{Mapping}, we
proceed with solving the problem. It turns out that this problem
is much more tractable than the conventional localization.
The localization transition driven by long-range hopping has
been studied previously and is well understood\cite{LL}. It is
known that as the decay rate of hopping changes, one comes to a
point at which there appears an infinite number of resonances,
occurring at all length scales. The interaction of these
resonances delocalizes the quantum states of the system.
However, right at the transition point the density of resonances
turns out to be low (provided that the coupling is weak), and
one can construct a theory of the transition by treating the
resonances as a kind of dilute gas. This makes it possible to
construct a renormalization group theory (RG). We discuss the
basics of the RG formalism in Sec.~\ref{DivergingResonances},
identify the quantities characterizing the strength of coupling,
and then in Sec.~\ref{RG} derive an RG flow of the coupling
parameters.

The RG equation derived in Sec.~\ref{RG} has some interesting
features. It has properties very similar to that of the Boltzmann
equation, and can be studied by tools borrowed from statistical
mechanics. In Sec.~\ref{Solving RG} we find integrals of motion
of the RG flow, derive an $H-$theorem, and obtain stationary
distributions characterizing universal RG limit.

The similarity between our RG equation and the Boltzmann equation has the
following physical meaning. The RG procedure involves subsequent
diagonalization of resonances ordered in energies, which is the
same as the ordering in increasing spatial scales, i.e., in the
RG time $\xi=\ln R$. Because of weak coupling, the only
important kind of resonances are pair resonances which are well
separated in the RG time domain. This situation is analogous to
the statistical mechanics of a gas, where because of low density
only pair collisions are important. Because of long collision
time, one can neglect the correlations of
subsequent collisions. Thus, everything about gas dynamics can
be extracted from the mechanics of a two-atom collision, which
is the essence of the Boltzmann equation. Similarly, in our problem
the role of elementary collision events is played by pair
resonances. The RG equation accounts for the change of the
coupling parameters resulting from one resonance, and treats the
resonances as uncorrelated events occurring randomly in the RG
time. This analogy is quite fruitful: one can define an
``entropy'' obeying an $H-$theorem, and show that the RG flow
leads to some analog of statistical equilibrium, in which the
distribution of couplings takes a universal limiting form (which
is the analog of Maxwell distribution in a gas).

In Sec.~\ref{P-ratio} we use the RG to study scaling properties
of the states. For that, the RG is extended to include
participation ratios of the states. We write down an RG equation
for the flow of the participation ratio distribution, and by
solving it determine the scaling exponent. The exponent turns
out to scale with the coupling strength, $Ze^2/\hbar v$, and
thus is small at weak coupling. This exponent has a geometric
meaning of fractal dimension of the space region occupied by
an eigenstate.

In Sec.~\ref{Transport} we study transport in the momentum
space, which in terms of Bloch states means angular diffusion.
As it should be at the localization transition, the diffusion is
of anomalous character, $t\sim {\cal D}^{-1} r^d$. We find
that the ``diffusion constant'' ${\cal D}$ scales with the coupling strength.
The dimensionless conductivity estimated from the Einstein relation $\sigma={\cal D}n$
is mush less than unity. Scaling
properties of the scale invariant density-density correlator are
related to the scaling exponent for the participation ratios
derived in Sec~\ref{P-ratio}.

Finally, let us mention that the renormalization method employed
in this work is not completely new. Similar ideas were used in
the studies of one-dimensional quasiperiodic Schr\"odinger
operators\cite{Thouless}, and of localization in
quasicrystals\cite{quasi-RG}.

\section{Model}\label{Model}

To be specific, we will focus on the problem of a particle moving
in a $3-$dimensional periodic crystal of scatterers:
     \be\label{crystal}
V(r) = \ {\sum_{a\in L}}\ U(r-a)\ ,
     \ee
where
$U(r)$ is a potential of one scatterer, and
$L$ is a $3-$dimensional lattice with periods $a_1$, $a_2$, and $a_3$:
  \be
a_1n_1{\bf i}+a_2n_2{\bf j}+a_3n_3{\bf k}\ , \qquad
{\rm where}\qquad
n_\alpha\in Z\ .
  \ee
The interaction $U(r)$ has a power law singularity at small $r$,
$U(r)\sim |r|^{-\sigma}$, and falls rapidly at distances much larger
than the lattice spacings $a_\alpha$, so that the
sum (\ref{crystal}) is well defined.  For the model we take
      \be\label{potential}
U(r) = \frac{\lambda}
{|r|^{\sigma}}\
e^{- \kappa |r|}
      \ee
where the ``effective radius'' $\kappa^{-1}$ of the interaction is
chosen to be of the order of the  distance  between the  scatterers:
$\kappa a_\alpha\simeq1$.

We are interested in the spectrum and eigenstates of the
Schr\"{o}dinger operator
   \be
-\frac{1}{2} \nabla^{2} + V(r)
   \ee
in the semiclassical range of energies:
      \be\label{energy_range}
E \gg  \frac {\hbar^{2}}{ma^{2}_\alpha}
      \ee
Eigenstates in the periodic potential are Bloch states $\psi_p(r)$
characterized by quasimomentum $p$ in the Brillouin zone of the
lattice $L$.

Physically, the model (\ref{crystal})--(\ref{energy_range})
corresponds to the problem of the scattering of a fast particle
by a crystal. Most interesting is the case of charged particles,
like electrons or muons, because it turns out that the Coulomb
potential singularity, $U(r) \sim \frac{1}{r}$ lies on a
critical line of a transition from regular to chaotic dynamics.
However, it is tutorial to consider the problem
(\ref{crystal})--(\ref{energy_range}) with an arbitrary
$\sigma$.

The problem (\ref{crystal})--(\ref{energy_range}) is closely
related to so-called quantum billiards. In the case of
billiards, typically, one considers a particle moving within a
rectangular, rhombic, or square billiard with a scattering
potential inside and a hard wall boundary condition on the
walls. In such billiards all nontrivial dynamics is due to the
scattering by the potential, since the dynamics in an empty
billiard is integrable. The problem of a hard core disk
potential is known as a Sinai billiard. It is natural to
generalize it to potentials with a power law singularity
(\ref{potential}).

The only difference between the billiard problem and our problem
is in boundary conditions: hard wall for billiards vs. periodic
in our case. It will be clear below that in both cases all
results are the same. In fact, in our discussion of the mapping
to a localization problem, as well as in the following treatment
of localization, one can everywhere replace Bloch waves by
standing waves without affecting any of the conclusions.
Clearly, this is consistent with the mathematical equivalence of
the billiard and the crystal problems that one has in the
semiclassical limit (\ref{energy_range}).

\section{Mapping to a localization problem}
\label{Mapping}

Let us write the problem in the basis of Bloch
plane waves $e^{i(p+g)r}$, where $p$ is the quasimomentum in the
Brillouin zone, and ${\bf g}$ is the vector of the lattice
$L^{\prime}$, reciprocal to $L$:
  \be\label{Gdef}
{\bf g}={2\pi\over a_1}n_1{\bf i}+{2\pi\over a_2}n_2{\bf j}+
{2\pi\over a_3}n_3{\bf k}\ , \qquad
{\rm where}\qquad
n_\alpha\in Z\ .
  \ee
In this basis Schr\"{o}dinger equation takes the form
   \be\label{4}
E  c_{\bf g}=E_{\bf g}c_{\bf g} + \sum_{\bf g^{\prime} \neq g}
U_{\bf g-{\bf g}^{\prime}} c_{\bf g^{\prime}},
   \ee
where
   \be\label{5}
E_{\bf g} = \frac {(p+g)^{2}}{2m} +  U_{0},
   \ee
and $U_{\bf g}$ are the potential Fourier components $U_{\bf g} =
V^{-1} \int e^{igr} U(r) d^{3} r$, where $V=a_1a_2a_3$
is the volume of the lattice unit cell. For the interaction
(\ref{potential}) one has
  \be\label{hopping1}
U_{\bf g}=\frac{4\pi\lambda}{|{\bf g}|^{3-\sigma}V}\
\frac{\Gamma(2-\sigma)\ \sin\left((2-\sigma)\tan^{-1}q/\kappa\right)}
{\left(\kappa^2/g^2+1\right)^{(2-\sigma)/2}}\ ,
  \ee
which can be expanded at small and large ${\bf g}$, correspondingly:
  \be\label{hopping2}
U_{\bf g}=\cases{
A_\sigma |{\bf g}|^{-(3-\sigma)}V^{-1}, &$|{\bf g}|\gg\kappa$;\cr
B_\sigma \kappa^{-(3-\sigma)}V^{-1}, &$|{\bf g}|\ll\kappa$,\cr
}
  \ee
where $A_\sigma=4\pi\lambda \Gamma(2-\sigma)\ \sin(\pi\sigma/2)$,
$B_\sigma=4\pi\lambda\ \Gamma(3-\sigma)$.

It is natural to  think  of  the  problem  (\ref{4})  as   an
Anderson  localization  problem  defined  on  the  sites  of the
lattice $L^{\prime}$, where $E_{\bf g}$'s are on-site energies,  and
$U_{\bf g-{\bf g}^{\prime}}$  are  hopping  amplitudes. Such an analogy is
meaningful because the hopping is {\it effectively local}, since
$U_{\bf g}$ falls at large $|{\bf g}|$. In the problem (\ref{4})  we  will
find   a   delocalization   transition  at  $\sigma=1$  and  any
$\lambda$, no matter how weak.

Let us recall that in the conventional Anderson model with a
hopping which is weak and short range, all states are localized.
In our problem this is not so, because the hopping is actually
long range: $U_{\bf g}\sim|{\bf g}|^{-(3-\sigma)}$ at $|{\bf
g}|\gg\kappa$. In such a problem, even weak hopping delocalizes
the states, provided the exponent in the hopping power law is
less than the space dimension (see
Sec.~\ref{DivergingResonances} and Refs.~\cite{LL,LL1}). We will
argue that the delocalization occurs over a shell of sites that
have almost equal energies. In the momentum space, such sites
are close to a ``Fermi sphere'' which has dimension two (see
Fig.~\ref{MomentumSpace}). Thus, at small $\lambda$ we have two
phases:\\ (i) $3-\sigma>2$, localized states;\\ (ii)
$3-\sigma<2$, delocalized states.\\ Localization in momentum
space ($\sigma<1$) means that the wave function in real space is
a linear combination of a small number of plane waves, or
harmonics, i.e., it describes a quasiperiodic motion. On the
other hand, the states delocalized in momentum space
($\sigma>1$) correspond to a sum of many plane waves with almost
equal wavelengths, but nearly isotropic distribution of
wavevectors. In this case the dynamics is chaotic.

To  see why effective space dimension equals two, let us replace
$U_{\bf g-{\bf g}'}$ in (\ref{4}) by an infinitely long range hopping with
a constant amplitude
   \be\label{6}
W\equiv \max\limits_{{\rm all}\,{\bf g}}\{U_{\bf g}\}=
B_\sigma \kappa^{-(3-\sigma)}V^{-1}\ .
   \ee
This problem corresponds to scattering  by  a  $\delta-$function
potential, and to chaotic dynamics.

Upon making the hopping infinite range all hopping amplitudes
increase, and therefore the number of sites ${\bf g}$ over which the
states delocalize should also increase. However, for the new
problem the states can be easily written explicitly:
  \be
c^{(i)}_{\bf g}={W C^{(i)}\over E^{(i)}-E_{\bf g}} \ ,
  \ee
where $C^{(i)}$ is defined as the sum of all $c_{\bf g}$'s,
  \be
C^{(i)}=\sum\limits_{\bf g} c^{(i)}_{\bf g}\ ,
  \ee
and the eigenenergy $E^{(i)}$ is found from the equation
  \be
\sum\limits_{\bf g} {W\over E^{(i)}-E_{\bf g}}=1\ .
  \ee
From that, one finds that the distribution of momenta in each
state is isotropic, and that the sites ${\bf g}$ over which the
$i-$th state will spread are all contained within the shell
  \be
E^{(i)}-\gamma W<E_{\bf g}<E^{(i)}+\gamma W\ ,
  \ee
where $\gamma$ is of the order of one.
Clearly, the quantity $W$ just sets an upper bound on the delocalization
shell in the problem (\ref{4}).

We are interested in the situation when $W \ll E$, and hence the
shell is very thin compared to its  radius.  Therefore,  we  can
think of the shell as of an effective two dimensional ``physical
space''  in which our localization problem is defined. The density
of the on-site energies $E_{\bf g}$ can be calculated from the  sphere
$|p|=p_0=\sqrt{2mE}$  surface  area,  $4\pi  p_0^2$, and the free
particle density  of  states  $\nu(E)=mp_0V/2\pi^2\hbar^3$.  The
density  of  the  on-site  energies,  taken per unit area of the
shell and per unit energy, is
  \be\label{density_of_states}
n=Vmp_0^{-1}(2\pi\hbar)^{-3}\ .
  \ee

Another  similarity  with  the  localization problem  is  that  the
energies $E_{\bf g}$ can be treated as {\it quasirandom numbers}. To
see this,  one  notes  that  $E_{\bf g}$'s  are  given,  up  to a
constant, by squares of the lattice vectors lengths. However, in
a generic lattice $L$ all vectors have different lengths, except
for those related by a sign reversal ${\bf g}  \rightarrow  -{\bf g}$.  This
implies  different  distances  to  the  constant  energy  sphere
$|p|=p_{0}$, and thus no occasional coincidence among $E_{\bf g}$'s.
Thus, for now we assume the energies uncorrelated,  and  discuss
consequences of their correlation later.

To summarize, we argued that the problem
(\ref{4})--(\ref{hopping1}), from the point of view of the theory of
localization, is equivalent to the problem
  \be \label{A-Equivalent}
Ec_{\bf g}=E_{\bf g}c_{\bf g}+\sum\limits_{\bf g'\ne g}U_{\bf g-{\bf
g}'}c_{\bf g'}\ .
  \ee
In the problem (\ref{A-Equivalent}) the sites ${\bf g}$ are  uniformly
distributed
over a plane, and have uniform density
$n=Vmp_0^{-1}(2\pi\hbar)^{-3}$. It is assumed that in the
semiclassical limit the number of lattice points around the constant
energy shell becomes so large that one can replace the finite size
sphere by an infinite plane without affecting localization properties. The
nature of
this approximation is similar to the one made in the conventional localization
theory when one goes from a finite system of size much bigger than the
localization radius
to an infinite system.

 The energies $E_{\bf g}$ in (\ref{A-Equivalent}) are quasirandom numbers
having unit spectral density, and the hopping amplitudes $U_{\bf g-{\bf g}'}$
are defined by
  \be
U_{\bf g-{\bf g}'}=\lambda_{new}|{\bf g}-{\bf g}'|^{-\alpha}\ ,
  \ee
where
  \be
\lambda_{new}=4\pi\lambda_{old}V^{-1} \Gamma(2-\sigma)\sin(\pi\sigma/2)\qquad
{\rm and}\qquad
\alpha=3-\sigma\ .
  \ee
For convenience of notation, below we ignore the rescaling
factor relating $\lambda_{new}$ and $\lambda_{old}$, and denote
$\lambda_{new}$ by $\lambda$. This will not cause any ambiguity
because from now $\lambda_{old}$ will not appear.

Also, in the calculation done below
we assume Coulomb potential for the scatterers, i.e., everywhere
set $\sigma=1$ and $\alpha=2$. The generalization to other
potentials will be discussed at the end of the paper, in
Sec.~\ref{Discussion}.

Finally, let us mention that it is possible to generalize our problem
to other space dimensions. In fact, all arguments on the localization vs.
delocalization behavior given above remain valid if one replaces the space
dimension $3$ by any ${d}$. In this case, the constant energy shell in
momentum space,
in which the localization problem is defined, has dimension ${d}-1$. The
Fourier
transform of a power law potential $U(r)=\lambda r^{-\sigma}$ goes
as $U_{\bf g}=\lambda |{\bf g}|^{{d}-\sigma}$. Then, by the localization
criterion which requires
that the power law in the hopping spatial dependence equals the space
dimension,
we get that the transition occurs at $\sigma$ satisfying the condition
  \be
\alpha\ \equiv\ {d}-\sigma\ =\ {d}-1 \ .
  \ee
This means that the Coulomb potential ($\sigma=1$) has critical singularity
in all space dimensions (obviously, with ${d}=1$ excluded).

\section{Delocalization argument: diverging number of resonances}
\label{DivergingResonances}

We take advantage of the mapping to the localization problem, and from now on,
until Sec.~\ref{Channeling}, we deal with the localization problem
with long range hopping. For brevity, we will use $d$ to denote
the space dimension in which localization takes place, which corresponds
to $d-1$ in the above sections. The hopping power law exponent
is $\alpha$ (equal to $d_{\rm old}-\sigma$). Let us
recall that the character of localization depends
on the nature of the hopping. For a short-range hopping the
transition occurs at a particular ratio of the hopping to the
on-site random potential. In contrast, for sufficiently
long-range hopping, all the states are delocalized, no matter how
weak the hopping is\cite{Anderson}. More precisely, in a
$ {d}-$dimensional systems with the $|{\bf g}|^{-\alpha} $ asymptotics of
hopping localization can exist only for
$ \alpha>{d} $ (see~\cite{LL}). If $ \alpha<{d} $, then localization
is destroyed, even at weak hopping. Thus, a
localization-delocalization transition occurs at $ \alpha={d} $.

As a side remark, many real systems of interest do fall on the
critical line $ \alpha={d}=3 $ (e.g., see~\cite{Burin}).
Among such systems are localized optical phonons in disordered
dielectric materials coupled by electric dipole forces,
two-level systems in glasses coupled by an $ r^{-3} $ elastic
interaction, magnetic impurities in metals coupled by the $ r^{-3} $
RKKY interaction, etc.

Let us briefly recall here the qualitative argument of Ref.~\cite{LL} showing
that the states of the problem (\ref{4})--(\ref{hopping1}) cannot be localized.

Consider two localized states located at ${\bf g}$ and ${\bf g}'$
having energies $E_{\bf g}$ and $E_{\bf g'}$.
They are {\em resonance\/} if
  \begin{equation}
|U_{\bf g-{\bf g}'}| \geq |E_{\bf g}-E_{\bf g'}|\qquad \label{(2.7)}
  \end{equation}
If the resonance condition (\ref{(2.7)}) holds, then eigenstates
of the problem
  \ber
Ec_{\bf g}=E_{\bf g}c_{\bf g}+U_{\bf g-{\bf g}'}c_{\bf g'}\cr
Ec_{\bf g'} =E_{\bf g'}c_{\bf g'}+U_{\bf g'-{\bf g}}c_{\bf g}
  \eer
are not localized on ${\bf g}$ or ${\bf g}'$, but are essentially non-zero
at both sites, ${\bf g}$ and ${\bf g}'$.

Now let us estimate the number of resonances and show that it
diverges at $\alpha={d}$. Let us takes one site ${\bf g}$, and
consider all sites ${\bf g}'$ which fall in resonance with the site ${\bf g}$.
The mean number of such sites is given by
  \be\label{ave_resonances}
\kappa_{d} \int |{\bf g}'|^{{d}-1}d|{\bf g}'|
P[{\bf g}, {\bf g'}]\ ,
  \ee
where $P[{\bf g}, {\bf g'}]$ is the probability that the condition
(\ref{(2.7)}) holds, and $\kappa_{d}$ is the surface area of
a ${d}-$dimensional unit sphere. To estimate $P[{\bf g}, {\bf g'}]$, we
note that since the on-site energies $E_{\bf g}$ are uniformly
distributed with the density $n$, one simply has
  \be\label{mean_probability}
P[{\bf g}, {\bf g'}]=|U_{\bf g-{\bf g}'}|n\ .
  \ee
Substituting the probability (\ref{mean_probability}) with
$U_{\bf g-{\bf g}'}\sim \lambda |{\bf g}-{\bf g}'|^{-\alpha}$ into
Eq.~(\ref{ave_resonances}),
one gets an estimate for the average number of resonances:
  \be\label{alpha_d}
\lambda n
\kappa_{d}\ \int |{\bf g}'|^{{d}-\alpha-1}d|{\bf g}'|\ .
  \ee
This expression converges at large $|{\bf g}'|$
when $\alpha >{d}$, and diverges
when $\alpha\le{\bf d}$. The divergence
means that as one looks further away from a given site,
there are more and more resonances. Obviously, in this case all
the states must be delocalized.

It is peculiar that the localization criterion does not involve
the strength of hopping, but only the decay rate at large
distances. In the standard localization problem, with a short
range hopping, the localization transition can be reached by
varying the hopping strength. Whereas in our problem, starting from the
localized state, the transition can be reached either by
increasing the hopping strength or by decreasing $\alpha$.
This opens up a possibility to study the localization transition at
hopping amplitudes much smaller than the on-site potential variance.
In this regime, due to relative smallness of the hopping, one can
develop an accurate theory of the transition  (see
Secs.~\ref{RG},~\ref{Solving RG}).

To see more clearly what kind of simplification one has in the weak
hopping limit, let us look again at Eq.~(\ref{alpha_d}) and
estimate the number of resonances at critical $\alpha={d}$.
One has
  \be\label{log-divergence}
\kappa_{d} n\ \int |{\bf g}'|^{-1}d|{\bf g}'| = \lambda n
\kappa_{d}\ \ln (R_{\rm max}/R_{\rm min})\ ,
  \ee
where $R_{\rm max}$ is the size of the system, and $R_{\rm min}$
is a ``microscopic'' scale. The number of resonances
(\ref{log-divergence}) increases logarithmically with the
system size, which indicates delocalization. However, if the hopping
strength $\lambda$ is small, the effect of delocalization will
be {\em weak}, because the resonances will occur rarely in the
$log-$space. Another way to express it is to note that
Eq.~(\ref{log-divergence}) means that the resonances have a uniform
density over the $log-$shells:
  \be
2^k R_{\rm min}< |{\bf g}-{\bf g}'| <
2^{k+1}R_{\rm min}\ , \qquad k=1,2,3,...
  \ee
This suggests treating the resonances ``shell by shell,''
i.~e., a renormalization group approach.
Since the density of the resonances distribution over
the shells is small in $\lambda$
one can construct an accurate and self-consistent RG scheme.

\section{Renormalization group theory.}
\label{RG}

In this section we review the RG theory of resonances\cite{LL}.
The reason an RG treatment is required in this problem is that
when all {\it direct} resonances discussed above are combined
together and diagonalized, there appear subsequent resonances,
basically of the same nature as primary resonances. Then, after
the next generation of resonances is treated, there appear
more resonances and this process must be repeated
(as shown schematically in Fig.~\ref{Hierarchy}).

To derive an RG equation an approximation will be made in which
only pair resonances are taken into account, while all higher
order resonances (triplets, quadruplets, etc.) are ignored. To
motivate this approximation, let us discuss one property of
resonances that will be basic for our approach.

Let two sites ${\bf g}$ and ${\bf g}'$ form a  resonance.  Consider  another
site  ${\bf g}''$  which  is  in  a  resonance  with any of these two.
According  to  Eq.~(\ref{log-divergence})  we  can  estimate   the
probabilities
  \begin{equation}
P\left( \frac{1}{2} \leq \frac{|{\bf g}'-{\bf g}|}{|{\bf g}''-{\bf g}|}
\leq 2 \right) \simeq \lambda n\ll 1\qquad
\label{(3.1)}
  \end{equation}
  \begin{equation}
P\left( \frac{\underline{1}}{2} \leq \frac{|{\bf g}''-{\bf g}'|}
{|{\bf g}-{\bf g}''|} \leq 2 \right) \simeq \lambda n \ll 1
  \end{equation}
(here 2 can be replaced by any other number of the order of one).
In other words, if the three sites ${\bf g}$, ${\bf g}'$, and ${\bf g}''$ fall
in a resonance, then one side of the triangle they form (say,
$|{\bf g}-{\bf g}''|$) is much shorter than the other two
($|{\bf g}-{\bf g}''|$ and
$|{\bf g}'-{\bf g}''|$). In accordance with Eq.~(\ref{log-divergence}), one
has
  \begin{equation}
\log _{2}\left( \frac{\min \left(|{\bf g}''-{\bf g}|,
|{\bf g}''-{\bf g}'|\right)}{|{\bf g}-{\bf g}'|} \right)
\simeq (\lambda n)^{-1} \gg 1\qquad
   \label{(3.2)}
  \end{equation}
Therefore, one concludes that pair resonances are much
more frequent than triplets: the estimate (\ref{(3.1)})
shows that the probability of a triplet resonance is of order of
$ \lambda^{2} $. Similar arguments show that the probabilities
of finding resonances of $ k $ oscillators $ \left(k=4,5,6,\dots
\right) $ can be estimated as $ \lambda^{k-1} $. This should be
compared with the probability of a pair resonance which is of
order of $ \lambda $ (see (\ref{(3.1)}) ).
We see that pair resonances occur about $\lambda^{-1} $ times
more frequently than triplets, about $\lambda^{-2} $ times more
frequently than quadruplets, and so on. This justifies keeping only
pairs in the RG.

The RG procedure involves the following steps. We truncate the
$|{\bf g}|^{-{d}}-$interaction at some $R_{0}$, i.e., put
$U_{\bf g-{\bf g}'}=0$ for all pairs $\left({\bf g}, {\bf g'}\right) $ such
that $
|{\bf g}-{\bf g}'|>R_{0} $. Let us assume we know exact eigenstates for the
truncated hamiltonian (call them $ R_{0}-$states). Then, to
eliminate next renormalization shell, we replace $ R_{0} $ by $
R_{1} $ such that
  \begin{equation}
R_{1} \gg R_{0}\ ,
\qquad \text{ but }\qquad
\lambda n \log _{2 }^{ } \frac{R_{1 }}{R_{0} } \ll 1\qquad
   \label{(3.3)}
  \end{equation}
Let us take the $ R_{1}-$states and represent them as linear
combinations of the $ R_{0}-$states. According to the above
discussion, all of $ R_{1}-$states are either single
$R_{0}-$states or resonance pairs of $ R_{0}-$states (we neglect
by triple and all higher order resonances).

The consistency of the RG procedure relies on the fact that the
localization radius of the $R-$states is less than $R$. One
notes that by going from $R_0-$states to $R_1-$states,
at weak hopping, the localization radius cannot exceed $R_1$. In
other words, on the scale $R$ the states will look like a
superposition of several states localized on a scale much
smaller than $R$.

Practically, this means that at all $R$ one can work in the
basis of localized states. In the RG scheme,
when considering the $R_{1}-$states as a result of
interaction (resonance) of the $R_{0}-$states we assume that the
resonating $R_{0}-$states are {\em far apart\/} compared to their
localization radii. This enables to keep track of the coupling
between the states, ignoring the details of the spatial
structure of wavefunctions.

The RG flow preserves the form of the Hamiltonian, but modifies the
hopping amplitudes. We will see below that the change of the amplitude $U_{\bf
g-{\bf g}'}$
of hopping between ${\bf g}$ and ${\bf g}'$ is given by a multiplier that
factors into
$\bar a_{\bf g'}a_{\bf g}$. The quantities $a_{\bf g}$ have a meaning of
coupling parameters
generated by the RG flow.
Mathematically, at each $R$ the truncated Hamiltonian has the form
  \be
E  c_{\bf g}=E_{\bf g}c_{\bf g}+\sum\limits_{|{\bf g'}-{\bf g}|<R}
U_{\bf g-{\bf g}'}\bar a_{\bf g} a_{\bf g'}\ c_{\bf g'}\ ,
  \ee
where the quasirandom energies $E_{\bf g}$ and the hopping
$U_{\bf g-{\bf g}'}=\lambda/|{\bf g}-{\bf g}'|^{\alpha}$ are the same as
in the original problem (\ref{4})--(\ref{hopping1}), and the new
quantities are the parameters $a_{\bf g}$ characterizing the strength
of coupling between the states. Initially, all $a_{\bf g}=1$. However,
the RG procedure generates a non-trivial distribution of the
parameters $a_{\bf g}$.

To see this in more detail, let us write down all necessary
quantities for two states (at $R=R_{0}$) located at ${\bf g}$ and
${\bf g}'$. They interact according to
  \ber \label{(3.4)}
E c_{\bf g}=E_{\bf g}c_{\bf g}+ U_{\bf g-{\bf g}'}\bar a_{\bf g} a_{\bf g'}\
c_{\bf g'}\ , \cr
E c_{\bf g'}=E_{\bf g'}c_{\bf g'}+ U_{\bf g'-{\bf g}}\bar a_{\bf g'} a_{\bf
g}\ c_{\bf g}\ ,
  \eer
Two eigenstates $ c^{+}$ and $c^{-}$ are defined by
  \begin{equation}
c^{+} = \cos \theta\ c_{\bf g} + \sin \theta\ e^{i\phi}\ c_{\bf g'}, \qquad
c^{-} = -\sin \theta\ c_{\bf g} + \cos \theta\ e^{i\phi}\ c_{\bf g'}\ ,
  \end{equation}
with
  \begin{equation} \label{(3.5)}
\tan\theta = \sqrt{\tau^2+1} - \tau\ , \qquad
\tau = \frac{E_{\bf g}-E_{\bf g'}}
{2|U_{\bf g-{\bf g}'}\bar a_{\bf g} a_{\bf g'}|}\ ,\qquad
\phi={\rm arg}(\bar a_{\bf g} a_{\bf g'})\ .
  \end{equation}
The energies of the states (\ref{(3.5)}) are
  \begin{equation} \label{(3.6)}
E_{\pm} = \frac{1}{2}\left( E_{\bf g}+E_{\bf g'}
\pm \sqrt{(E_{\bf g}-E_{\bf g'})^2 +4|U_{\bf g-{\bf g}'}\bar a_{\bf g} a_{\bf
g'}|^2}\right)
  \end{equation}
The transformation rule for the parameters $a_{\bf g}$ can be found from the
relation
  \begin{equation} \label{(3.7)}
a_{\bf g} c_{\bf g}+a_{\bf g'} c_{\bf g'} = {a}^{+}c^{+} + {a}^{-}c^{-}\ ,
  \end{equation}
which means that
  \begin{equation}
{a}^{+}= \cos \theta\ {a}_{\bf g} + \sin \theta\ e^{-i\phi}\ {a}_{\bf g'}\ ,\qquad
{a}^{-}= -\sin \theta\ {a}_{\bf g} + \cos \theta\ e^{-i\phi}\ {a}_{\bf g'}\qquad
  \label{(3.8)}
  \end{equation}
The transformation rule (\ref{(3.8)}) for the coupling
parameters $a_{\bf g}$ shows that even when initially all $a_{\bf g}=1$ the
resonances lead to non-unit values of $a_{\bf g}$, which justifies
including $a_{\bf g}$'s in the RG.

As a matter of fact, the parameters $a_{\bf g}$ are the only important
ingredient of the RG. Other characteristics of the states, such
as their energies, turn out to be irrelevant. The reason is that
when a resonance of two states forms (say, at the scale $R$) the
energies of the states split by something like $|U_{\bf g-{\bf g}'}|\sim
R^{-\alpha}$. (Since at the scale $R$ the resonances are
typically spaced by $|{\bf g}-{\bf g}'|\sim R$.) This splitting ensures that
at larger scales these states never fall in resonance. Because
of that, the energies $E_{\bf g}$ can be taken as uncorrelated
quasirandom numbers with uniform distribution, and their change
under RG transformations can be ignored.

Now, let us proceed with deriving an RG equation for the
distribution of $a_{\bf g}$'s. In doing it we emphasize the analogy
with the theory of Boltzmann kinetic equation. Boltzmann
equation for gases is usually derived assuming the absence of
correlations of subsequent collision processes, resulting from
large mean free paths of molecules in a gas. Besides providing
grounds for a probabilistic approach, the largeness of mean free
path enables one not to take into account triple and other
multiple collisions. Similarly, in our RG dynamics there is no
repetition of resonances (collisions) of the same states,
because of the energy splitting. Also, small density of the
resonances in the $log-$space, analogous to large mean free path
in gases, makes triple and higher order resonances unimportant.

As a first step of deriving the RG equation, let us consider how
the distribution of $a_{\bf g}$'s changes by going from $R_0-$states
to $R_1-$states. The distribution $f(a)$ is defined by the
probability density
  \be
dP=f(a)dad\bar a\ .
  \ee
Under the change $R_0\to R_1$ the distribution $f(a)$ changes to
$\widetilde{f}(a)$. The change is due to resonances
(``collisions'') occurring during the RG time interval $\ln R_0 <
\xi < \ln R_1$. The difference
$\widetilde{f}\left({a}\right)-f\left({a}\right) $ can
be obtained by integrating
  \begin{equation}
\frac{1}{2} \left[ \delta\left({a}-{a}^{+}\right) +
\delta\left({a}-{a}^{-}\right) - \delta\left({a}-
{a}_{1}\right) - \delta\left({a}-{a}_{2}\right) \right]\qquad
   \label{(3.9)}
  \end{equation}
over
  \\
{\it 1)} $ f\left({a}_{1}\right)d{a}_{1}d\bar a_1$;
  \\
{\it 2)} $ f\left({a}_{2}\right)d{a}_{2}d\bar a_2$;
  \\
{\it 3)} $nd^{2}{\bf g}$ , where ${\bf g}={\bf g}_1-{\bf g}_2$ and
$ n$
is the concentration of sites given by Eq.~(\ref{density_of_states}).
  \\
{\it 4)} $dE$ , where $E=E_1-E_2$.

Here the subscripts ``1'',~``2'' label the states in a resonance
pair. It is convenient to introduce a new variable $\tau$
instead of $ E $, defined by the second equation in
(\ref{(3.5)}). The rotation angle $ \theta $ is related with $
\tau $ by $\tan\theta=\sqrt{\tau^{2}+1}-\tau $. The usefulness
of the variable $\tau$ becomes clear from the identity
  \ber
d^{2}{\bf g} dE &=& 2\pi |{\bf g}| d|{\bf g}|\ 2|\bar a_1 a_2 U_{\bf g_1-{\bf
g}_2}|\ d\tau =
4\pi |{\bf g}|^{2}|\bar a_1 a_2 U_{\bf g_1-{\bf g}_2}|\ |{\bf g}|^{-1}d|{\bf
g}| d\tau \cr
&=& \left(4\pi |{\bf g}_{1}-{\bf g}_{2}|^{2}|\bar a_1 a_2 U_{\bf g_1-{\bf
g}_2}|\right)
d\left(\ln |{\bf g}|\right)
d\tau\qquad \label{(3.11)}
  \eer
where the product $|{\bf g}_{1}-{\bf g}_{2}|^{2}|\bar a_1 a_2 U_{\bf g_1-{\bf
g}_2}|$ does not
depend on
$|{\bf g}_{1}-{\bf g}_{2}|$. (Because
$U_{\bf g_1-{\bf g}_2}=\lambda |{\bf g}_{1}-{\bf g}_{2}|^{-2}$.)
Thus, we perform an integration over
${\bf g}_1-{\bf g}_2 $ in the domain $R_0<|{\bf g}_1-{\bf g}_2|<R_1$. The result is
  \ber
\widetilde{f}\left({a}\right)-f\left({a}\right) =
\ln \left(R_{1}/R_{0}\right) \lambda n \int d\tau
d\bar a_1 d{a}_{1}
d\bar a_2 d{a}_{2}
\ |a_1|\ |a_2|\
f\left({a}_{1}\right)
f\left({a}_{2}\right) \cr
\times
\Bigl[\delta\left({a}-
{a}^{+}\right)+
\delta\left({a}-{a}^{-}\right)-\delta\left({a}-
{a}_{1}\right)-\delta\left({a}-{a}_{2}\right)\Bigr]
\qquad
   \label{(3.12)}
  \eer
According to the above discussion of the probability of pair
resonances, the RHS in (\ref{(3.12)}) is of order of $ \lambda n
\ln \left(R_{1}/R_{0}\right) $, which is a small number (see
(\ref{(3.3)}) ). Then it is standard to replace
Eq.~(\ref{(3.12)}) by a differential equation with respect to
the RG time $\xi=\ln \left(R\right)$:
  \ber
\frac{\partial}{\partial\xi}f\left({a}\right) &=& \lambda n \int d\tau
d\bar a_1 d{a}_{1}
d\bar a_2 d{a}_{2}
\ |a_1|\ |a_2| \cr
&\times &
f\left({a}_{1}\right) f\left({a}_{2}\right)
\Bigl[\delta\left({a}-
{a}^{+}\right)+\delta\left({a}-{a}^{-}\right)-
\delta\left({a}-{a}_{1}\right)-\delta\left({a}-
{a}_{2}\right)\Bigr]\ .
\label{(3.13)}
  \eer
This equation is the main result of this section. Our task now
will be to look for solutions $f\left({a},\xi\right)$ to the
problem (\ref{(3.13)}) with the $\delta-$function initial condition $
f\left({a},\xi=0\right) =\delta(a-1)$ corresponding to the
``microscopic'' distribution of the parameters ${a}_{i}$.

\section{Solving the RG equation}
\label{Solving RG}

The similarity with the Boltzmann equation makes the
dynamics (\ref{(3.13)}) very simple. The
problem (\ref{(3.13)})
has integrals of motion
(analogous to the conservation of energy and momentum for
Boltzmann equation), and also it obeys an $H-$theorem. Relying
on these properties we analyze the asymptotical behavior of $
f\left({a},\xi\right) $ at $ \xi\to\infty $.

First, let us study the integrals of the dynamics
(\ref{(3.13)}). The simplest integral is
 \be
\langle |a|^2\rangle=\int d\bar a da |a|^2 f(a,\xi )
  \ee
Let us prove that
  \begin{equation}
\frac{\partial}{\partial\xi}\langle |a|^{2}\rangle =0\qquad
\text{ or }\qquad
\int|a|^{2}f\left({a},\xi\right)d\bar{a}da = \int|a|^{2}f\left({a},0\right)
d\bar{a}da
  \label{(4.1)}
  \end{equation}

\noindent
{\em Proof:\/}
  \begin{equation}
\frac{\partial}{\partial\xi}\langle |a|^{2}\rangle =n \int d\tau\int
d\bar{a}_{1}da_1
d\bar{a}_{2}da_2
f\left({a}_{1}\right)f\left(
{a}_{2}\right)\left[\left|{a}^{+}\right|^{2}+\left|
{a}^{-}\right|^{2}-\left|{a}_{1}\right|^{2}
-\left|{a}_{2}\right|^{2}\right]\quad
    \label{(4.2)}
  \end{equation}
But
  \be
\left|{a}^{+}\right|^{2}+\left|{a}^{-}\right|^{2}=
\left|\cos \theta\ {a}_{1}+\sin \theta\ e^{-i\phi}\ {a}_{2}\right|^{2}+
\left|-\sin \theta\ {a}_{1}+\cos \theta\ e^{-i\phi}\ {a}_{2}\right|^{2}=
\left|{a}_{1}\right|^{2}+\left|{a}_{2}\right|^{2}
  \ee
(unitary transformations preserve the norm). Consequently,
the RHS of Eq.~(\ref{(4.2)}) vanishes.\qquad {\it QED}

Besides $\langle|a|^{2}\rangle$ there exist other invariants of
Eq.~(\ref{(3.13)}). All averages of the real and imaginary parts
of $a=a'+ia''$,
  \be
\langle a'^{2}\rangle\ ,\qquad
\langle a''^{2}\rangle\ ,\qquad
\langle a'a''\rangle\ ,
   \label{(4.3)}
  \ee
are conserved by the same argument of orthogonal transformations
preserving the norm.

An interesting issue to discuss is whether the conservation of
the quantities (\ref{(4.3)}) is an exact or an approximate
result. One might worry that the conservation fails when
higher order resonances are taken into account,
in addition to the interacting pairs.
However, the conservation laws remarkably survive.
To see why that is, let us consider a resonance of $k$ states. The Hamiltonian (\ref{4}) is
diagonalized by an orthogonal $k\times k$ matrix
$U_{\alpha\beta}$, ($\alpha,\beta =1,\dots ,k$). One notes that
the parameters $a_\alpha$ are transformed by the the matrix
$U^{-1}$, and then the conservation of the quantities
(\ref{(4.3)}) follows from orthogonality of $U$.

Another property of the problem
(\ref{(3.13)}) is an analog of the $H-$theorem.
Let us define {\it entropy} of the distribution $f(a,\xi)$ as
  \begin{equation}
H_\xi \left[f\right] = - \int \ln \left(|{a}|f\left({a},\xi\right)
\right) f\left({a},\xi\right) d\bar{a}da
  \label{(4.5)}
  \end{equation}
The $H-$theorem states that $H_\xi\left[f\right]$ grows monotoneously
as function of the RG time $\xi$:
  \begin{equation}
\frac{\partial}{\partial\xi} H_\xi\left[f\right] \geq 0\ .
  \label{(4.6)}
  \end{equation}
{\em Proof:\/} Using the identity $ 0 = \frac{\partial}{\partial\xi} 1 =
\frac{\partial}{\partial\xi}\int f\left({a},\xi\right)d\bar{a}da =
\int\frac{\partial}{\partial\xi}f\left({a},\xi\right)d\bar{a}da $
one finds
  \ber
\frac{\partial H}{\partial\xi} &=& -\int\ln \left(|{a}|f\right)
\frac{\partial f}{\partial\xi}d\bar{a}da    \cr
&=& n \int d\tau\int\int d\bar{a}_{1}da_1 d\bar{a}_{2}da_2 |{a}_{1}
|f\left({a}_{1}\right)
|{a}_{2}|f\left({a}_{2}\right) \ln \left[
\frac{|{a}_{1}|f\left({a}_{1}\right)|{a}_{2}
|f\left({a}_{2}\right)}{|{a}^{+}|f\left({a}^{+}\right)
|{a}^{-}|f\left({a}^{-}\right)}\right]
   \label{(4.7)}
  \eer
Another useful identity
  \be\label{usefulidentity}
\int d\tau\left[\int\int|{a}_{1}|f\left({a}_{1}\right)
|{a}_{2}|f\left({a}_{2}\right)d\bar{a}_{1}da_1
d\bar{a}_{2}da_2\
-\int\int|{a}^{+}|f\left({a}^{+}\right)
|{a}^{-}|f\left({a}^{-}\right)d\bar{a}^{+}da^{+}
d\bar{a}^{-}da^{-}
\right]=0
  \ee
follows from the fact that the variables ${a}_{1}$, ${a}_{2} $ and
${a}^{+}$, ${a}^{-} $ are related by an
orthogonal transformation. Combining the identity (\ref{usefulidentity})
with (\ref{(4.7)}) one gets
  \begin{equation}
\frac{\partial H}{\partial\xi} = n \int d\tau\int\int d\bar{a}_{1}da_1
d\bar{a}_{2}da_2
\Phi\left({a}^{+}\right)\Phi
\left({a}^{-}\right)\left(X\ln(X)-X+1\right)\text{ ,\qquad }
   \label{(4.8)}
  \end{equation}
where $ X = (\Phi\left({a}_{1}\right)\Phi\left({a}_{2}
\right))/(\Phi\left({a}^{+}\right)\Phi\left({a}^{-}\right))$, and
$\Phi\left({u}\right)=|{u}|f\left({u}\right) $.

Finally, the quantity $X \ln(X) - X + 1$ is non-negative. From that,
the $H-$theorem
(\ref{(4.6)}) follows momentarilly.\qquad {\it QED}

Note that although our entropy differs from Boltzmann entropy by
$\langle \ln|a|\rangle$, it does not affect the proof
of the $H-$theorem which follows the standard derivation (cf.
Ref.~\cite{Boltzmann}).

The problem (\ref{(3.13)}) has {\it stationary distributions}
given by
  \begin{equation}
f_{G}\left({a}\right) = \frac{A}{|{a}|}\exp \left(-a_{\alpha}
G_{\alpha\beta}a_{\beta}\right)\text{ ,\qquad }
   \label{(4.10)}
  \end{equation}
where $a_1=a'$, $a_2=a''$, and $ \widehat{G} $ is a positively
defined symmetric $ 2\times 2$ matrix.

\noindent
{\em Proof:\/} If a solution $ f\left({a},\xi\right) $ of Eq.~(\ref{(3.13)})
does not depend on $ \xi
$ then $ \frac{\partial H}{\partial\xi}=0 $ for it.
Since
$X\ln(X)-X+1=0 $ vanishes only for $ X=1 $, according to
Eq.~(\ref{(4.8)}) for such a solution
$\Phi\left({a}_{1}\right)\Phi\left({a}_{2}\right)=
\Phi\left({a}^{+}\right)\Phi\left({a}^{-}\right)
$ for all $ \theta $ (hence, for all $ \tau $). This is possible provided that
  \begin{equation}
f\left(a\right) = f_{G}\left(a\right)\text{ .\qquad }
   \label{(4.11)}
  \end{equation}
Given $G_{\alpha\beta}$, the normalization constant $A$ in (\ref{(4.10)})
should be determined from
the condition $ \int f\left({a}\right)d\bar{a}da=1 $.
\qquad {\it QED}

Stationary solutions $ f_{G}\left({a}\right) $ are analogous to the
Maxwell distribution which appears in statistical mechanics as a
stationary solution of Boltzmann equation.
An important distinction is that in the Maxwell distribution there is only
one free parameter (temperature) while the distributions
$f_{G}\left({a}\right) $ are characterized by 3 parameters
$G_{\alpha\beta}$ ($\alpha,\beta=1,2$; $\alpha\leq\beta $). Like in
the Boltzmann equation theory, the number of parameters in the
stationary distribution is equal to the number of conserved
quantities (\ref{(4.3)}).

Continuing the analogy with statistical mechanics, let us prove
that our entropy is maximal on the stationary distributions
(\ref{(4.10)}). More precisely, let us take all functions
$f\left({a}\right) $ such that
  \begin{equation}
\int f\left({a}\right)d\bar{a}da=1\text\ ,\qquad
\int a_{\alpha}a_{\beta}f\left({a}\right)d\bar{a}da
=G_{\alpha\beta}\ , \qquad
\left(\alpha,\beta=1,2\right)\ ,\qquad
   \label{(4.12)}
  \end{equation}
where $a_1=a'$, $a_2=a''$, $a=a'+ia''$.
Then always $ H\left[f\right]\leq H\left[f_{G}\right] $, where
$f_{G}\left({a}\right) $ is defined by (\ref{(4.10)});
$H\left[f\right]=H\left[f_{G}\right] $ only if
$f\left({a}\right)=f_{G}\left({a} \right) $.

\noindent
{\em Proof:\/} Let us consider
the first and the second variations of $H[f]$ constrained by (\ref{(4.12)})
and by  the normalization condition:
  \ber
\delta\left[ H\left[f\right]-\lambda\int f\left({a}\right)d\bar{a}da-
\sum_{\alpha\beta}\mu_{\alpha\beta}\int a_{\alpha}a_{\beta}f\left({a}
\right)d\bar{a}da \right] \cr
= \int\left[-\ln \left(|{a}|f\left({a}
\right)\right)-1
-\lambda-\sum_{\alpha\beta}\mu_{\alpha\beta}a_{\alpha}a_{\beta} \right]
\delta f\left({a}\right) d\bar{a}da
  \eer
  \begin{equation}
\delta^{2}\left[ H\left[f\right]-\lambda\int f\left({a}\right)
d\bar{a}da -\sum_{\alpha\beta}\mu_{\alpha\beta}\int a_{\alpha}a_{\beta}f
\left({a}\right)d\bar{a}da \right] = -\int \frac{\left(\delta f
\left({a}\right)\right)^
{2}}{f\left({a}\right)} d\bar{a}da  <\text{ 0.}
  \end{equation}
Since the second variation is negative, maximal value of $ H\left[f\right]
$ corresponds to
  \begin{equation}
f\left({a}\right) = \frac{1}{|{a}|}\exp
\left(-\left(1+\lambda\right)-\sum_{\alpha\beta}
\mu_{\alpha\beta}a_{\alpha}a_{\beta}\right)\qquad
   \label{(4.13)}
  \end{equation}
The relationship between $  \lambda,\mu_{\alpha\beta}  $  and  $
A,G_{\alpha\beta}  $  is straightforward (compare (\ref{(4.13)})
with (\ref{(4.10)}) ).\qquad {\it QED}

The asymptotic properties of solutions of Eq.~(\ref{(3.13)})
are thus rather simple: each solution converges to one of the stationary
distributions $ f_{G}\left({a}\right) $. The parameters $ G_{\alpha\beta} $
are completely determined by second moments of the initial distribution
$ f\left({a},\xi=0\right) $:
  \begin{equation}
G_{\alpha\beta}=\int a_{\alpha}a_{\beta}f\left({a},\xi=0\right)d\bar{a}da
\left(\alpha,\beta=1,2, \alpha<\beta\right).
  \end{equation}

To conclude, the RG dynamics (\ref{(3.13)}) has a
$3-$dimensional family of fix points parameterized by symmetric
positively defined $2\times2$ matrices. The dynamics in the space of distributions
$f(a)$ is shown scematically in Fig.~\ref{DynamicsView}.

The parameters are
determined by the initial distribution, which means that the
coupling strength in this problem is not renormalized. Thus the
critical behavior such as scaling of the states will depend on
these parameters. This situation should be contrasted with the
ordinary localization transition at short-range hopping, where
the critical behavior is believed to be completely universal,
with no additional parameters.

Let us mention that for the particular problem of Coulomb scatterer we are
interested in the distribution $f_G(a)$ has the form
  \be\label{F-real}
f(a')=A|a'|^{-1}\exp\left(-Aa'^2\right)\ ,
  \ee
i.e., it does not
depend on the imaginary part $a''$. The reason is that in the initial
distribution all parameters $a$ are real:
$f(a)|_{\xi=0}=\delta^{(2)}(a-1)$.
The non-generic form of matrix $G_{\alpha\beta}$ resulting from
the fact that all $a$'s are real leads to a
complication: the stationary distribution (\ref{F-real})
is not normalizable. However, since the divergence of $\int f(a') da'$ is only
logarithmic (i.e., weak), the results in this case are the same as for generic $G_{\alpha\beta}$.
The only subtlety is that the constant $A$ in (\ref{F-real}) is renormalized, at
a rate proportional to $\ln\xi$, which leads to the distribution slowly flowing
to more and more singular distributions. However, all results derive above still
hold, including the absence of RG correcions to the coupling strength $\langle (a')^2\rangle$.
Because of that, the conclusion about non-universal RG fix point holds in this case as well.

\section{Wavefunction scaling: RG for the distribution of participation
ratios}
\label{P-ratio}

In this section we study renormalization of the eigenstates. The
hierarchical structure of the states (resonances of resonances
of resonances ...) emerging from the RG dynamics suggests that
they have fractal properties. To study the scaling behavior of the states
we will consider their participation ratios. For a normalized
state $c^{(i)}_{\bf g}$ its participation ratio $p^{(i)}$ is defined
by
  \be
p^{(i)}=\sum\limits_{\bf g} |c^{(i)}_{\bf g}|^{4}
  \ee
In the theory of localization participation ratios are used as a
measure of the degree of localization. By studying scaling of
participation ratios one can characterize fractal properties of
the states.

Below we study how participation ratios are changed by the real
space RG of the states described in Sec.~\ref{RG}. For this
purpose we need participation ratios $ p_{R}^{(i)}$ of the
$R-$states, which have been defined as true eigenstates of the
system with the interaction truncated at the scale $R$. To
incorporate the participation ratios in the RG, one can rewrite
the RG formalism for the distribution $f(a,p)$, including both
coupling strengths and participation ratios. In fact, the
$R-$states are characterized not only by $a^{(i)}$ and
$p^{(i)}$, but also by eigenenergies and positions. However, as
we already discussed, the correlations between the energies and
positions are not important, and thus at each step of the RG one
can assume uniform uncorrelated distribution over energy and in
space.

As it was argued above,
the RG flow
in the leading order is determined by pair resonances. Let us discuss how pair resonances affect
the participation ratios, and derive an RG flow for the
distribution $f(a,p)$. For that we consider the resonance pair
(\ref{(3.4)}) and find for it the participation ratios $ p^{\pm}
$ of the states ``$+$'' and ``$-$'' as functions of $p_{i}$ and
$p_{j}$. Since the positions of the states $i$ and $j$ are well
separated, one simply has:
  \begin{equation}
p^{+} =\ \cos^{4}\theta\  p_{i} + \sin^{4}\theta\  p_{j},\qquad
\qquad p^{-} =\ \sin^{4}\theta\  p_{i} + \cos^{4}\theta\  p_{j}.
    \label{(7.3)}
   \end{equation}
This change of the participation ratios must be considered
together with the transformation
of the parameters $a_i$, $a_j$ given by
Eq.~(\ref{(3.8)}). Other steps of the derivation of the RG flow
for $ f\left({a},p\right) $ are similar to those described in
Sec.~\ref{RG}, and the resulting RG equation is the following:
  \ber
\frac{\partial}{\partial\xi}f\left({a},p\right)
&=&
n \int d\tau d\bar {a}_{1}da_1 d\bar {a}_{2}da_2
dp_{1}dp_{2}\ |a_1|\ |a_2|\
f\left({a}_{1},p_{1}\right)f\left({a}_{2},p_{2}\right) \cr
&\times& \big[\delta\left({a}-{a}^{+}\right)\delta\left(p-p^{+}\right)+\delta
\left({a}-{a}^{-}\right)\delta\left(p-p^{-}\right)  \cr
&-&
\delta\left({a}-{a}_{1}\right)\delta\left(p-p_{1}\right)-\delta\left({a}-{a}_{
2}\right)\delta
\left(p-p_{2}\right)\big]\  .\qquad
    \label{(7.4)}
   \eer
Of course, from the flow equation (\ref{(7.4)}) one can return
to Eq.~(\ref{(3.13)}) by using the relation between the
functions $ f\left({a}\right) $ and $ f\left({a},p\right) $:
  \begin{equation}
f\left({a}\right) = \int_{0}^{\infty} f\left({a},p\right) dp\qquad
   \label{(7.2)}
   \end{equation}

By using the RG flow for the distribution $f(a,p)$ given by
Eq.~(\ref{(7.4)}) one can study the $\xi-$dependence of various
moments of $p$ and of $a$, such as $\langle p^{k}\rangle$ or
$\langle|a|^{m} \rangle$, as well as correlators $\langle
p^{k}|a|^m \rangle$. Here we concentrate only on the
distribution of the participation ratios, assuming the
distribution of the parameters $a$ to be stationary. In other
words, we will be interested only in the scaling limit of the
problem.

Let us focus on calculating the average participation ratio
  \begin{equation}
\langle p\rangle = \int\int\int d\bar {a}da dp\ p
f\left({a},p\right)|_{\xi=\ln R}\ .
   \label{(7.5)}
   \end{equation}
To facilitate the analysis we introduce a new function
$s\left({a}\right) $ defined by
   \begin{equation}
s\left({a}\right) = \int_{0}^{\infty} p f\left({a},p\right) dp\ .
   \label{(7.6)}
   \end{equation}
The RG flow for $s(a)$ can be obtained
by integrating Eq.~(\ref{(7.4)}) over $p$:
  \ber
\frac{\partial}{\partial\xi}s\left({a}\right) =
n \int d\tau d\bar {a}_{1}da_1 d\bar {a}_{2}da_2
dp_{1}dp_{2}\ |a_1|\ |a_2|\
f\left({a}_{1},p_{1}\right)f\left({a}_{2},p_{2}\right) \cr
\times\big[ p^{+}\delta\left({a}-{a}^{+}\right)+p^{-}
\delta\left({a}-{a}^{-}\right)-
p_{1}\delta\left({a}-{a}_{1}\right)-p_{2}\delta\left({a}-{a}_{2}\right)\big]
\ .
   \label{(7.7)}
   \eer
The integrals over $ p_{1}$ and $p_{2} $ in Eq.~(\ref{(7.7)}) can be
calculated by
using (\ref{(7.6)}), (\ref{(7.3)}) and (\ref{(7.2)}):
  \begin{equation}
\frac{\partial}{\partial\xi}s\left({a}\right) = n \int d\tau d\bar {a}_{1}da_1
d\bar{a}_{2}da_2
\ |a_1|\ |a_2|\
\Big[s\left({a}_{1}\right)f\left({a}_{2}
\right)\left[..\right]_{1}
+ s\left({a}_{2}\right)f\left({a}_{1}\right)\left[..\right]_{2 }\Big] \ ,
   \label{(7.8)}
   \end{equation}
where
  \begin{equation}
\left[..\right]_{1}=\ \cos^{4}\theta\
\delta\left({a}-{a}^{+}\right)+\sin^{4}\theta\
\delta\left({a}-{a}^{-}\right)-\delta\left({a}-{a}_{1}\right)
   \end{equation}
and
  \begin{equation}
\left[..\right]_{2}=\ \sin^{4}\theta\
\delta\left({a}-{a}^{+}\right)+\cos^{4}\theta\
\delta\left({a}-{a}^{-}\right)-\delta\left({a}-{a}_{2}\right)
   \label{(7.9)}
   \end{equation}
In the scaling limit $\xi\to \infty$, as we demonstrated
above, the distribution $ f\left({a}\right) $ converges at
large $\xi$ to a stationary distribution
$f_{G}\left({a}\right)$. Since we are interested in the
asymptotical behavior of $ s\left({a}\right) $ at $
\left(\xi\to\infty\right)$, the function $ f\left({a}\right) $
in Eq.~(\ref{(7.8)}) can be set equal to the stationary
distribution $f_{G}\left({a}\right) $ given by Eq.~(\ref{(4.10)}).
After that we get a {\em linear\/} integral equation for $
s\left({a}\right) $ with a kernel idependent of
$\xi $.

With that in mind, one can view  Eq.~(\ref{(7.8)}) as
  \begin{equation}
\frac{\partial}{\partial\xi}s\left({a}\right) =
-\widehat{B}\left(s\left({a}\right)
\right)\ ,
   \label{(7.10)}
   \end{equation}
where $ \widehat{B} $ is a constant ($\xi-$independent) linear operator
which acts in the space of functions
$s\left({a}\right) $. Asymptotical behavior of $ s\left({a}
\right) $ at $ \xi\to\infty $ is then given by
  \begin{equation}
s\left({a},\xi\right) \simeq \exp \left(-\mu\xi\right)
s_{0}\left({a}\right)\ ,
   \label{(7.11)}
   \end{equation}
where $ \mu $ is the {\it lowest eigenvalue} of $ \widehat{B} $ and $
s_{0}\left({a}\right) $ is the
corresponding eigenvector: $ \widehat{B}\left(s_{0}\left({a}\right)\right)=\mu
s_{0}\left({a}
\right) $. Then, according to
Eq.~(\ref{(7.5)}), the scaling behavior of $\langle p\rangle$ is given by
  \begin{equation}
\langle p\rangle
= \int d\bar {a}da  s\left({a}\right)
|_{\xi= \ln R} \simeq R^{-\mu}\ .
   \label{(7.12)}
   \end{equation}
Therefore, the scaling exponent of $\langle p\rangle$
is related to $\mu$, the lowest eigenvalue of $ \widehat{B} $.

In principle, if the stationary distribution $f_{G}\left({a}\right) $ is
known, $ \mu $ can be determined (perhaps,
numerically). However, here we are interested in estimating
$\mu$ only by the order of magnitude.

To estimate $\mu$, let us note that the kernel
in Eq.~(\ref{(7.8)}) is of the order of
  \begin{equation}
n \langle |{a}_{1} {a}_{2}| \rangle \simeq n \langle {a}^{2}\rangle
=\lambda n\ .
   \label{(7.14)}
   \end{equation}
Hence, given $n$ and $\lambda$,
  \begin{equation}
\mu = \gamma \lambda n\ ,
   \label{(7.15)}
   \end{equation}
where $\gamma$ is a constant of the order of one.

It is tutorial to derive the result (\ref{(7.15)}) by another
method. Using Eq.~(\ref{(7.8)}) one can write
  \ber
-\mu &=& \frac{\partial}{\partial\xi} \ln \left(\langle p\rangle
\right)|_{\xi\to\infty}
= \langle p\rangle^{-1}\
\int \frac{\partial}{\partial\xi}s\left({a}\right)d\bar {a}da  \cr
&=& n \langle p\rangle^{-1}\
\int d\tau
d\bar{a}_{1}da_1 d\bar {a}_{2}da_2\ |{a}_{1}|\ |{a}_{2}|\ \Bigl[
s\left({a}_{1}\right)
f_{G}\left({a}_{2}\right)\left[..\right]_{1} +
s\left({a}_{2}\right)f_{G}\left({a}_{1}
\right)\left[..\right]_{2 } \Bigr]
\ , \label{(7.16)}
  \eer
where
  \begin{equation}
\left[..\right]_{1}= \left[..\right]_{2} = \sin^{4}\theta +\cos^{4}\theta -1.
\qquad \label{(7.17)}
   \end{equation}
But $ \sin^{4}\theta +\cos^{4}\theta -1 = -2\sin^{2}\theta  \cos^{2}\theta  <
0$, and hence Eq.~(\ref{(7.16)}) can be rewritten as:
  \be\label{(7.16a)}
-\mu = -4n\ \int d\tau \sin^2\theta\cos^2\theta \ \int d\bar a da f_G(a)\
\left(\int d\bar a da |a| s(a) \left/
\left(\int d\bar a da s(a)\right)\right.\right)
  \ee
By estimating the RHS of Eq.~(\ref{(7.16a)})
by the order of magnitude, one again obtains the result (\ref{(7.15)}) and in
addition proves that $\mu$ is positive.

The exponent $\mu$ has a simple geometric meaning. For a
typical $R-$state, it gives the fractal dimension of a region
occupied by this state. Hence, we find that in this problem
the fractal
dimension of the states depends on the
coupling strength $\lambda n$.

The argument presented above for the mean participation ratio,
after proper modification, is also valid for any moment of the
participation ratios. Hence, all other fractal dimensions must scale
with $\lambda n$ the same way as $\langle p\rangle$.

In the localization problem it is sometimes of interest to
consider the distribution of all moments of the wavefunction:
  \be
p(m)=\sum\limits_{\bf g} |c^{(i)}_{\bf g}|^{m}\ .
  \ee
The participation ratio studied above corresponds to $m=4$. The
distribution of $p(m)$ can be used to study fractal
dimensions of the states.

It is straightforward to generalize our RG formalism for the quantities
$p(m)$. For that, one simply notes that for the pair resonance
the change of $p(m)$'s is written as
  \be
p^{+}(m) =|\cos\theta |^m\  p_{i}(m)+ |\sin\theta |^m\ p_{j}(m) ,\qquad
\qquad p^{-}(m)=|\sin\theta |^m\ p_{i}(m)+|\cos\theta |^m\ p_{j}(m)\ .
    \label{p(n)}
  \ee
Putting it together with the transformation (\ref{(3.8)}) leads
to the RG flow for the distribution $f(p(m),a)$ of the same
form as Eq.~(\ref{(7.4)}), but with $p^{\pm}$ given by
Eq.~(\ref{p(n)}). The analysis of this RG equation is similar to
that presented above, and so are the results. The scaling
exponents for the quantities $p(m)$ are of the order $\lambda n$.
This confirms our conclusion about fractal dimension of the
space region occupied by an eigenstate being of the order of
$\lambda n$.

\section{Transport: Anomalous diffusion and scaling}
\label{Transport}

In this section we study time-dependent transport.
Particularly we are interested in the density correlation
function
  \begin{equation}
K\left({\bf g}-{\bf g}', t\right) =
\sum\limits_{i, j} \
\langle e^{-i(E_i-E_j)t} \
\bar c^{(i)}_{\bf g} c^{(j)}_{\bf g}
\bar c^{(j)}_{\bf g'} c^{(i)}_{\bf g'}
\rangle \ ,
  \label{(5.1)}
  \end{equation}
where $c^{(i)}_{\bf g}$ are eigenstates, $E_i$ are their energies, and
${\langle}\dots {\rangle}$ stands for averaging over the on-site
disorder.

We will argue that the dynamics is critical (anomalous
diffusion) and write down a scale invariant expression
(\ref{(5.7)}) for the correlation function $K(g, g', t)$. Then,
by using the RG for the participation ratios presented in
Sec.~\ref{P-ratio} we will derive scaling properties of the
function $K$.

Let us begin with a qualitative picture of propagation of an
excitation in the system. Let the particle start at $t=0$ from
one of the sites (say, from the site 1 in Fig.~\ref{Hierarchy}).
After some time $ T $ it hops to its nearest resonance
neighbor, site 2. The hopping time $T$ is of the order of inverse
resonance splitting, which, according to Sec.~\ref{DivergingResonances},
can be estimated as
  \be
T\sim |E_{+}-E_{-}|^{-1}\approx
|{\bf r}_{1}-{\bf r}_{2}|^{2}/\lambda\ .
  \ee
After the time of the order of $T$ the particle is localized not
only at the site $ {\bf r}_{1} $, but both at $ {\bf r}_{1} $ and
at $ {\bf r}_{2} $, since both modes $c^{+}$ and $c^{-}$ are
excited.

Later, after some larger time $ T'\gg T$ new site will be involved
in transport. Each of the two excited states sets in
motion other states which themselves can be considered as a
result of interaction of several states during the time
interval $ \leq T' $. This is shown in Fig.~\ref{Hierarchy},
where new resonance states
participating in the transport are linear
combinations of $c_{3}$, $c_{4} $ and of $ c_{5}$, $c_{6} $,
respectively. Obviously,
$ L'=|{\bf r}_{1}-{\bf r}_{\alpha}|\simeq|{\bf r}_{2}-{\bf r}_{\alpha}|\gg
|{\bf r}_{1}-{\bf r}_{2}|$, where $\alpha=3,4,5,6 $.
Thus, at times $t$ which are longer than
$T'$ but
shorter than the time necessary for exciting new states,
the excitation is mainly carried by four states.
According to the estimates made in Sec.~\ref{DivergingResonances}, one has
  \begin{equation}
L'/L\simeq|{\bf r}_{1}-{\bf r}_{\alpha}|/|{\bf r}_{1}-{\bf r}_{2}|\simeq
\exp \left({\rm const}/(\lambda n)\right)\ ,\qquad T'/T\simeq
\left(L'/L\right)^{2},\quad \label{(5.3)}
  \end{equation}
where $ \alpha=3,4,5,6 $.

Even later, after some time $ T'' $
$ \left(T''\gg T'\right) $ new
states are involved in transport. The corresponding
sites are separated from the initial site
by a large distance $ L'' $ such that
  \begin{equation}
L''\text{{\em/\/}}L'\simeq\exp \left({\rm const}.\lambda^{-1}\right)
\ ,\qquad T''\text{{\em/\/}}T'\simeq\left(L''\text{{\em/\/}}L'\right)^{2}\ .
   \label{(5.4)}
  \end{equation}
At this stage, the number of modes carrying the excitation again
approximately doubles.

Let us remark here that the excitation evolution described above
should be understood in a statistical sense. For a particular
state, there is no exact length scale at which the number of
resonances doubles. As we saw in Sec.~\ref{DivergingResonances},
the hierarchy of resonances is characterized by density uniform
in the log of spatial scale. The same is true for the time
hierarchy of resonances.

To summarize, the larger the time, the more states are excited.
Asymptotically, after a large time $ T $ about
  \begin{equation}
2^{c\lambda n\log \left(T\right)} = T^{c\lambda n}\qquad
  \label{(5.5)}
  \end{equation}
states are involved in the transport (here $ c $ is a number of the order of
one
to be determined later).

There are two conclusions one can draw. First, the RG picture of
the states hierarchy in space is in a one--to--one
correspondence with the hierarchy in the time domain. The
sequence of times at which new states are involved in
dynamics is determined by the spatial hierarchy of resonances.

Second, according to the above discussion, after a long time $T$
the excitation spreads over a region of size
$L\left(T\right)\simeq T^{1/2} $. This means that
the dynamics is
(asymptotically) {\it invariant\/} under rescaling
  \begin{equation}
\left({\bf r},t\right) \to \left(Z{\bf r},Z^{2}t\right)\ ,
  \label{(5.6)}
  \end{equation}
where the rescaling factor $ Z $ is an arbitrary number.

In particular, the correlation function $ K\left({\bf
r},t\right) $ must be invariant under the rescaling
(\ref{(5.6)}). Thus we have
  \begin{equation}
K\left({\bf r},t\right)= t^{-1} F\left(t^{-1/2}{\bf r}\right)
\ ,
  \label{(5.7)}
  \end{equation}
at large $t$, $|{\bf r}|$. The function $F({\bf x})$ is normalized,
  \be
\int F\left({\bf x}\right)d^{3}{\bf x}=1\ ,
  \ee
due to probability conservation. The particular form of $F({\bf x})$
depends in some universal way on the distribution of
couplings $f(a)$.

Let us note here that the simple scaling relation (\ref{(5.7)})
is a direct consequence of the existence of stationary fix
points $f_{G}(a)$ of the RG flow for $f(a)$ given by
Eq.~(\ref{(3.13)}). Indeed, the coupling of complex modes
composed of many single-site states is determined by the
distribution of the parameters ${a}_{i}$ given by the limiting
distribution $f_{G}\left({a}\right)=\lim_{\xi\to\infty}
f\left({a},\xi\right) $. Because this limit exists, the coupling
of complex $R-$states asymptotically scales as $R^{-2}$. The
time of hopping scales as inverse coupling, which leads to the
space-time scaling relation $T\sim R^2$.

There is a relation between the behavior of the scaling
function $F({\bf x})$ and small ${\bf x}$
and the participation ration scaling exponent
$\mu$ discussed in Sec.~\ref{Transport} (see
Eqs.~(\ref{(7.12)}),~(\ref{(7.15)})).
It turns out that
  \be\label{Fscaling}
F({\bf x})=|{\bf x}|^{-(2-\mu)} \qquad {\rm at}\qquad |{\bf x}|\ll 1\ .
  \ee
To prove this relation, let us consider a finite system of size $R$
and take the density correlator $K$ given by (\ref{(5.1)}) at times $t\le
R^2$.
By the order of magnitude, $K$ can be estimated by removing from
Eq.~(\ref{(5.1)})
all terms with non-equal energies $E_i\ne E_j$. (This is correct because
typically
$|E_i-E_j|$ is larger than level spacing, which is of the order $R^{-2}$, and
hence
the phase factors $(E_i-E_j)t$ in (\ref{(5.1)}) are large, which
leads to random relative signs of all
terms with $i\ne j$, and leads to mutual cancellations.)

On the other hand, after the terms with $i\ne j$ are
removed, one gets the following estimate:
  \be
K({\bf g}, {\bf g}',\ t\simeq R^2)\ \simeq\ \sum\limits_i
|c^{(i)}_{\bf g}|^2 |c^{(i)}_{{\bf g}'}|^2\ .
  \ee
Now, let us set ${\bf g}'={\bf g}$ and sum over all ${\bf g}$:
  \be \label{K(g=g')}
\sum\limits_{{\bf g'}={\bf g}}
K({\bf g}, {\bf g}',\ t\simeq R^2)\ =
\ \sum\limits_{i, {\bf g}}
|c^{(i)}_{\bf g}|^4\ =\ \sum\limits_{i} p^{(i)}\ ,
   \ee
where $p^{(i)}$ are the participation ratios. The right hand side of
Eq.~(\ref{K(g=g')}) is of the order
$R^{2-\mu}$ (see Eqs.~(\ref{(7.12)}),~(\ref{(7.15)})). This is consistent with
the scaling form
  \be
K({\bf g}, {\bf g}', t)=\ t^{-1}F(|{\bf g}-{\bf g}'|/t^{1/2})
  \ee
only if the function
$F$ satisfies Eq.~(\ref{Fscaling}).

Also, let us remark that in an arbitrary space dimension ${d}$ the
critical law of hopping,
$U_{\bf g -{\bf g}'}\sim |{\bf g}-{\bf g}'|^{{\bf -d}}$, leads to
anomalous diffusion: $T\sim R^{d}$. The reason is that, as it is
clear from the above discussion, the scaling transformation
(\ref{(5.6)}) changes to
  \begin{equation}
\left({\bf r},t\right) \to \left(Z{\bf r},Z^{d}t\right)\ ,
  \label{(5.9)}
  \end{equation}
and the scale-invariant function $F({\bf x})$ is now defined by
  \begin{equation}
K\left({\bf r},t\right)= t^{-1} F\left(t^{\bf -1/d}{\bf r}\right)
\ .
  \label{(5.10)}
  \end{equation}
($\int F\left({\bf x}\right)d^{d}{\bf x}=1$).
The anomalous diffusion follows directly from (\ref{(5.10)}):
  \begin{equation}
\langle |{\bf r}|^{2}\rangle = \int|{\bf r}|^{2}K\left({\bf
r},t\right)d^{d}{\bf r} \approx
t^{\bf 2/d}\int|{\bf x}|^{2}F\left({\bf x}\right)d^{d}{\bf x}\ .
  \label{(5.8)}
  \end{equation}
In the problem of three-dimensional Coulomb scatterers, the
space in which we have a localization problem is two-dimensional
(constant energy surface in momentum space), and thus in this
case there is no distinction between anomalous diffusion
and ordinary diffusion.

To compare the results with the standard localization theory
assuming short-range hopping,
let us point out that diffusion constant in owr problem is of
the order $\lambda$. Dimensionless conductivity can be found from the
Einstein relation $\sigma=n{\cal D}$ to be of the order $\lambda n\ \ll\ 1$.
Thus in this problem one has critical behavior with $\sigma\ll 1$.
On the other hand, in the delocalized phase of the system with short-range hopping
one always has $\sigma\le 1$.

\section{Scattering of charged particles in crystals}
\label{Channeling}

Let us return to the problem of charged particle scattering
in a crystal, and consider the dynamics of a fast charged partcile
(electron or muon) in a real crystal.
Since atoms cores are just $1/r-$scatterers,
one expects that the effects of weak chaos we
discussed above will take place, and lead to delocalization in
momentum space, i.~e., to the absence of Bloch waves. Of course,
from a practical view point, there is always some inelastic
scattering, which complicates observing the
effects we are interested in. However, in the following we ignore such
processes, as well as all possible effects of crystal boundaries.

Let us estimate the dimensionless coupling constant $\lambda n$.
From (\ref{density_of_states}) and (\ref{hopping2}) one has
  \be \label{Cconstant}
\lambda n = {4\pi Z e^2\over \hbar v}\ ,
  \ee
where $Z$ is the nucleus charge and $v$ is the particle
velocity. The coupling constant is of the order of one when the
velocity $v$ is of an atomic scale.

From (\ref{Cconstant}) it is clear that with electrons one never
has a strong coupling situation for a semiclassical problem. The
reason is that if an electron is moving with velocity of an
atomic scale, $v=Ze^2/\hbar$,
or slower, the number of wavelengths in the region
where the core potential is not screened is of the order of one.
On the other hand, if the velocity increases so that the
wavelength becomes small, one has a well defined, but not
a very interesting semiclassical problem, because the coupling
(\ref{Cconstant}) becomes very weak. Therefore, channeling of
fast electrons (or positrons) is not a suitable situation.

On the other hand, one gets into the desired regime much easier
with heavier particles, of which the best studied is channeling
of positive and negative muons. If a muon travels with velocity
of an atomic scale $Ze^2/\hbar$, so that the coupling
(\ref{Cconstant}) is of the order of one, its momentum is very
large:
  \be
p_{\mu} \simeq {m_{\mu}\over m_e} {\hbar\over a_B} \gg p_{atomic}={\hbar\over
a_B}
  \ee
Correspondingly,  the  particle  wavelength  will  be very small
compared to the  core  potential  radius  $a_B$,  and  thus  the
problem is semiclassical and our results are applicable.

The theory we described above predicts delocalization in
momentum space. In terms waves scattering it means that a plane
wave, after multiple scattering on the core potentials, will
give rise to many other plane waves, and eventually there will
be almost no memory preserved about initial plane wave. Following the
discussion of Sec.~\ref{DivergingResonances}, the number of
plane waves which are in a resonance with given plane wave can
be estimated as
  \be
\lambda n\ \ln \big( p_{\mu}/p_{atomic}\big) \gg 1
  \ee
At large $p_\mu/p_{atomic}$ one can have many resonances between
plane waves even for a weak coupling situation, $\lambda n \le
1$. In this case, the theory presented in this paper will be
accurate.

There are several simple predictions which can be tested
experimentally. It follows from our discussion of resonances
that the number of plane waves into which initial plane wave is
scattered, although large, is much smaller than the total number
of plane waves with energies close to the initial energy. The
origin of this effect is in the critical nature of delocalized
states for Coulomb potential. The diffusion over constant energy
surface in momentum space leads to a nontrivial angular
correlation function, with a power law singularity
(\ref{Fscaling}) at small angles.

\section{General picture}
\label{Discussion}

Our discussion of the localization in momentum space focused
on the Coulomb scatterers which have marginal singularity and
give rise to critical states in momentum space. It is easy to
see how the RG results will change for a general power law
singularity, $U(r)=\lambda/r^\sigma$. The hopping in momentum
space, given by Fourier transform of $U(r)$, will be more
short-range if $\sigma<1$ and less short-range if $\sigma>1$.
Hence, at $\sigma>1$ the system is always in the delocalized
phase, no matter what $\lambda$. On the other hand, for
$\sigma<1$ the RG will scale coupling down to zero, and we can
conclude that for sufficiently small $\lambda$ (i.e., when the
RG is applicable) all states are localized.

To represent this graphically, one can draw a schematic phase
diagram in the plane ($\sigma,\ \lambda$) (see Fig.~\ref{Diagram}).
The region marked by {\bf I} corresponds to the standard
delocalization transition for short-range hopping (large
$\sigma$). The region marked by {\bf II} corresponds to the
transition at small $\lambda$ studied in this article. One can
note that the critical behavior in the cases {\bf I} and {\bf
II} is quite different: all scaling relations in the case {\bf
II} depend on the dimensionless coupling strength which does not
change under the RG flow, whereas in the case {\bf I} there are no
parameters. This means that one cannot go continuously from one
scaling regime to the other, and hence we conjecture that there
must be a point on the localization-delocalization phase
boundary at which the critical behavior changes.

Another reason to believe that there is such a point is that in
the RG theory presented above the transition occurs at the same
$\sigma=1$ for any $\lambda$ (which must be sufficiently small).
So the phase boundary in Fig.~\ref{Diagram} has a part that goes
straight up. This suggests that the special point at which the
critical behavior changes must simultaneously be the point {\bf
C} where the phase boundary turns away from the $\sigma=1$
straight line.

One more reason supporting this phase diagram follows from
recent paper by Mirlin and Fyodorov (see Ref.~\cite{MirlinFyodorov}). They
studied a one-dimensional problem with power-law hopping in the
regime of strong hopping where one can use the supersymmetric
sigma-model. In the ($\sigma,\ \lambda$) plane this corresponds
to large $\lambda$. In this region, a delocalization transition
was found at certain critical $\sigma$. It is natural to make a
connection with our results by assuming that the point {\bf C}
is a tricritical point at which all phase boundaries match.

To summarize, in this article we consider the problem of Bloch
states for a fast particle moving in a crystal of scatterers
with power law singularity. This problem is shown to be
equivalent to a localization problem in momentum space which we
study using an RG previously developed for the localization with
long-range hopping. We find a localization transition with the
critical singularity being that of Coulomb potential. In the
localized phase the particle state is close to a single plane
wave, whereas in the delocalized phase the state is a
superposition of many plane waves with wavevectors spread
uniformly over a constant energy shell in momentum space. In
other words, the localization transition is the transition
between integrable and ergodic dynamics. We believe that the
relation between delocalization in momentum space and the
transition to chaos is very generic, and will be useful beyond
the problem of power-law scatterers. One can hypothesize that
understanding of Quantum Chaos in general is in some way a
problem of finding an equivalent in localization theory.

\begin{acknowledgements}
We are very grateful to Ilya Zakharevich for helping to generate
a LaTeX code from an unpublished manuscript written in 1989 on which
part of this article is based. The work of L.~L. was supported
partially by the MRSEC Program
of the National Science Foundation under award number DMR
94-00334.
   \end{acknowledgements}

\begin{center}
%
%
%
%
%
%

  \begin{figure}
\label{MomentumSpace}
\epsfig{file= 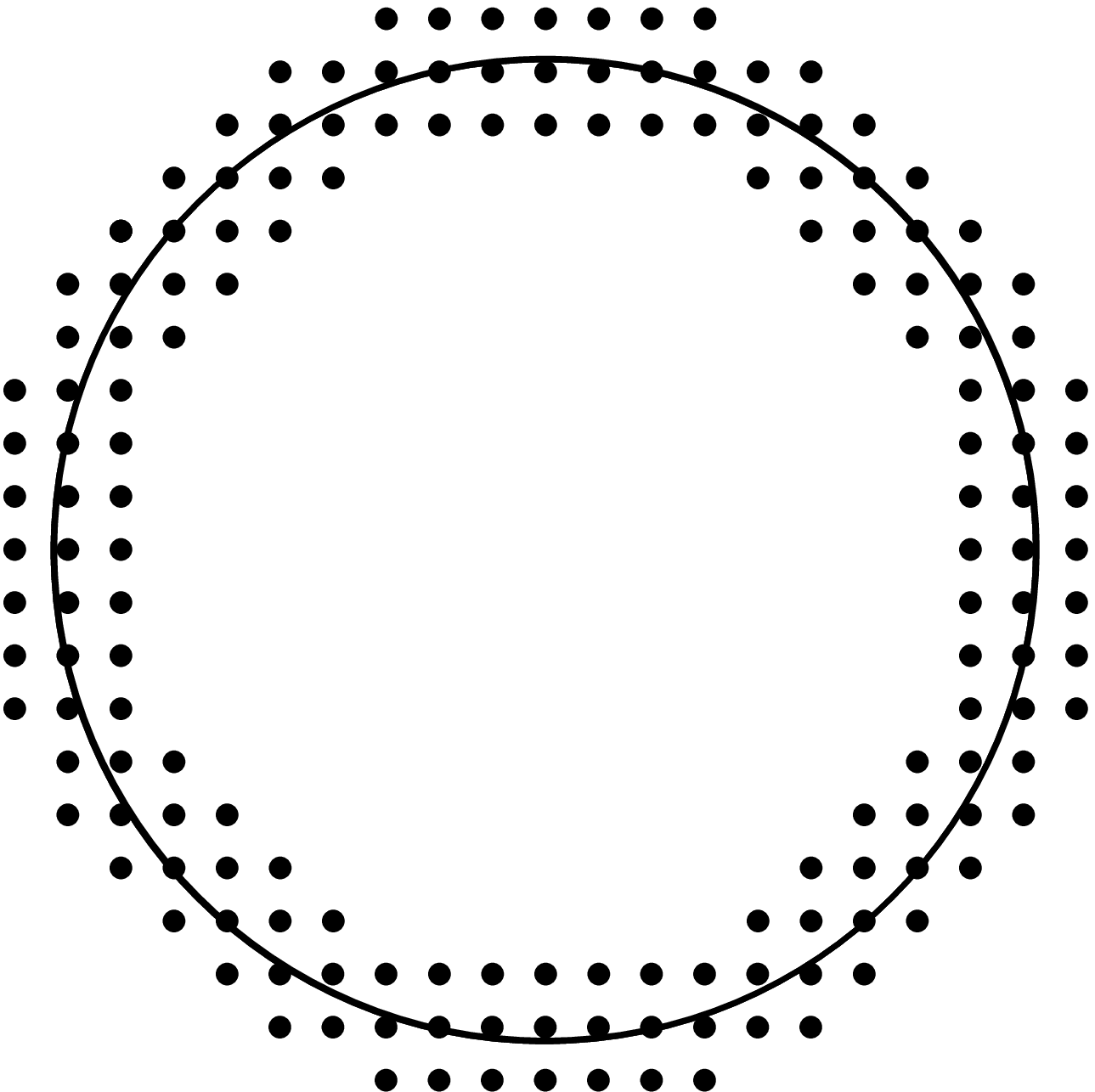, height= 90mm,  angle= 0}
\caption{
A sphere of constant energy is shown in momentum space, together
with a shell of the dual crystal lattice sites (\protect\ref{Gdef})
which are close to the sphere. Delocalization of Bloch states in
momentum space occurs within this shell. Effective dimension of
the space in which we consider the localization problem is given by the
sphere dimension, i.e., equals two.
    }
  \end{figure}

  \begin{figure}
\label{Hierarchy}
\epsfig{file= 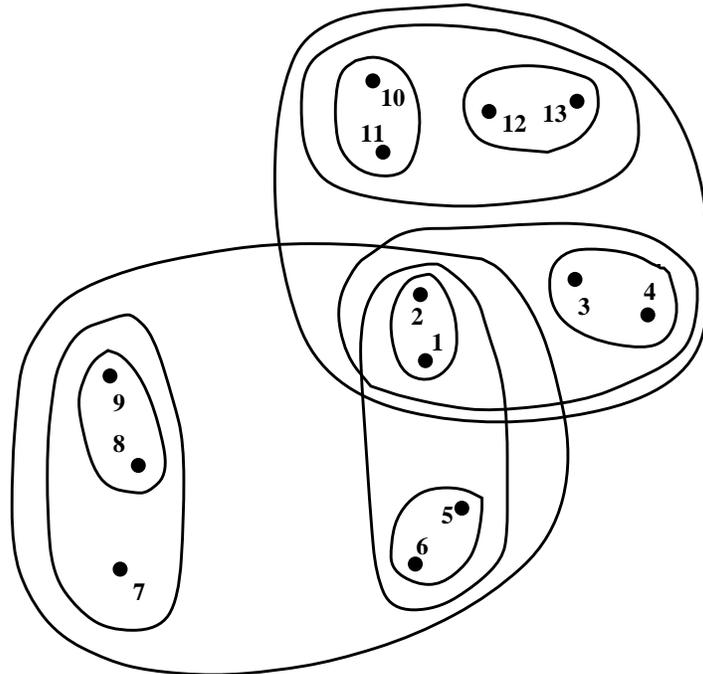, height= 90mm,  angle= 0}
\caption{
Hierarchical structure of two eigenstates is shown:\hskip5mm
    {\it a)} (((1+2)+(6+5))+((8+9)+ 7)) ;\hskip5mm
    {\it b)} (((1+2)+(3+4))+((10+11)+(12+13))) ;\hskip5mm
where we denote by $(n+m)$ a resonance pair formed by the states
$m$ and $n$. The time hierarchy of excitation transport is given by the
same diagram.
    }
  \end{figure}

  \begin{figure}
\label{DynamicsView}
\epsfig{file= 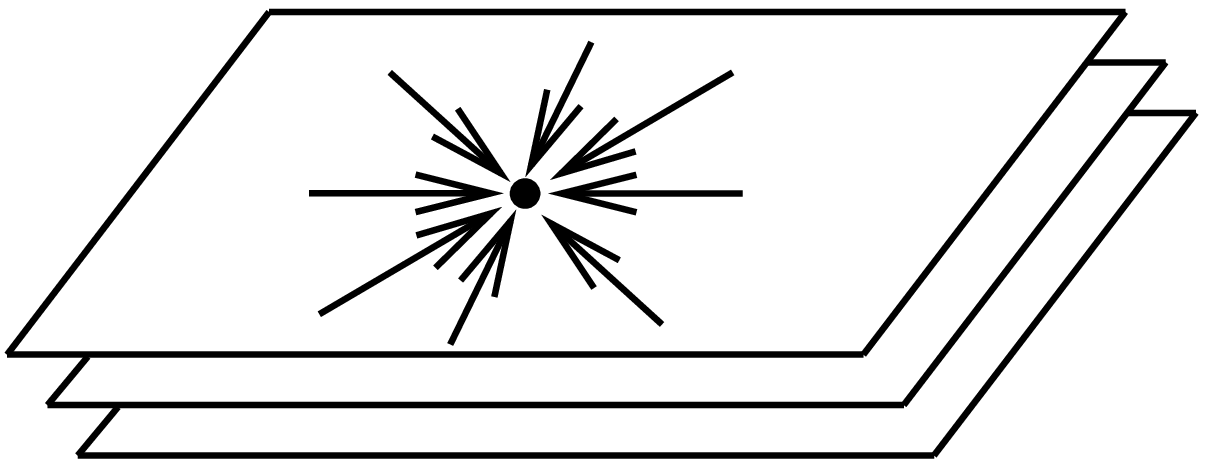, height= 90mm,  angle= 0}
\caption{
Schematic view of the RG flow given by Eq.~(\protect\ref{(3.13)}).
The space of functions $f(a)$ is represented
by a bundle of invariant surfaces labeled by the integrals
$G_{\alpha\beta}=\langle \bar a_{\alpha}a_{\beta}\rangle$,
($\alpha, \beta=1,2$, $\alpha\leq\beta$). Restricted on each of
the surfaces the system has one attracting fix point (given by
Expr.~(\protect\ref{(4.10)})).
    }
  \end{figure}

  \begin{figure}
\label{Diagram}
\epsfig{file= 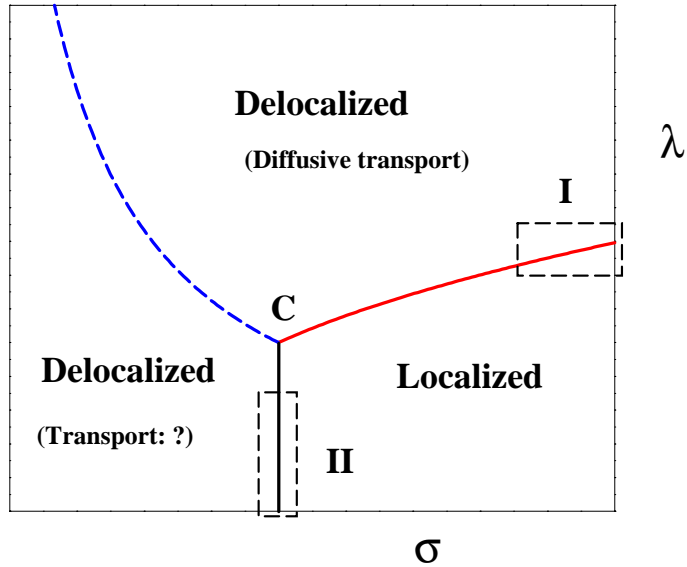, height= 90mm,  angle= 0}
\caption{
Phase diagram in the $(\sigma-\lambda)$ plane. (Both $\sigma$
and $\lambda$ vary between $0$ and $\infty$.) The phase
boundaries are marked corresponding to the delocalization by
long-range forces studied in this work (region {\it II}) and to
the conventional Anderson transition at short-range hopping
(region {\it I}).
    }
  \end{figure}

\end{center}

\end{document}